\documentclass[letterpaper, 10 pt, conference]{ieeeconf}

\IEEEoverridecommandlockouts                   \usepackage{nicematrix}

\usepackage[hidelinks]{hyperref} 
\usepackage{relsize}
\usepackage{amsmath,amssymb,amsfonts,mathtools}
\usepackage{nicematrix, nicefrac}
\usepackage{algorithmic}
\usepackage{textcomp}
\usepackage{xcolor}

\usepackage{booktabs}
\usepackage{graphicx}
\usepackage{subcaption}
\usepackage{epsfig} 
\usepackage{times}  
\usepackage{siunitx}
\NewCommandCopy\OldSI\SI
\RenewDocumentCommand\SI{ O{} m m }{%
  \nobreak%
  \OldSI[#1]{#2}{#3}%
  \nobreak%
}

\usepackage{tikz}
\usepackage{xstring}
\usepackage{forloop}
\usepackage{colortbl}
\usepackage{soul}
\usepackage{array}

\let\labelindent\relax  %
\usepackage{enumitem}

\usepackage{multirow} 
\usepackage{multicol}
\usepackage{arydshln}
\usepackage{pgfgantt}

\usepackage{textcomp}
\usepackage{rotating}
\usepackage{float}
\usepackage{algorithmic}
\usepackage{algorithm}
\usepackage[edges]{forest}

\usepackage[
  backend=biber,
  bibstyle=ieee,        %
  citestyle=numeric-comp, %
  sorting=none,         %
  sortcites=true        %
]{biblatex}

\usepackage{booktabs}

\newcommand{\g}{\textnormal{\textsl{g}}}

\renewcommand{\c}{\mathrm{c}}
\renewcommand{\d}{\nabla}

\renewcommand{\c}{{\mathrm{c}}}

\newcommand{\norm}[1]{\left\|#1\right\|}
\newcommand{\dotp}[2]{\langle #1,#2\rangle}

\newcommand{\cB}{{\mathcal B}}
\newcommand{\cC}{{\mathcal C}}

\newcommand{\cN}{{\mathcal N}}

\newcommand{\cW}{{\mathcal W}}
\newcommand{\cX}{{\mathcal X}}

\newcommand{\R}{\mathbb R}
\newcommand{\rS}{\mathbb S}
\newcommand{\col}[1]{\mathrm{col}\!\left(#1\right)}

\makeatletter
  \renewcommand*\env@matrix[1][*\c@MaxMatrixCols c]{%
	\hskip -\arraycolsep
	\let\@ifnextchar\new@ifnextchar
	\array{#1}}
\newcommand*\bigcdot{\mathpalette\bigcdot@{1.5}}
\newcommand*\bigcdot@[2]{\mathbin{\vcenter{\hbox{\scalebox{#2}{$\m@th#1\cdot$}}}}}
\makeatother

\newcounter{excounter}[section]

\newcounter{rmcounter}

\newcounter{dfcounter}

\newcounter{thmcounter}

\newcounter{corcounter}[thmcounter]

\newcounter{prcounter}

\newcounter{clcounter}

\graphicspath{{figures/}}
\usepackage[dvipsnames]{xcolor}

\usepackage{amsthm}
\usepackage{mathrsfs} 
\usepackage{placeins}
\newtheorem{theorem}{Theorem}
\newtheorem{lemma}[theorem]{Lemma}

\theoremstyle{definition}

\newtheorem{problem}{Problem}

\theoremstyle{definition}
\newtheorem{remark}{Remark}

\makeatletter

\newcommand{\Rmnum}[1]{\expandafter\@slowromancap\romannumeral #1@}
\makeatother

\AtEveryBibitem{\clearfield{doi}}
\AtEveryBibitem{\clearlist{organization}}
\AtEveryBibitem{\clearfield{month}}
\AtEveryBibitem{\clearfield{isbn}}
\AtEveryBibitem{\clearfield{issn}}
\AtEveryBibitem{\clearfield{series}}
\AtEveryBibitem{\clearfield{publisher}}
\AtEveryBibitem{\clearfield{url}}
\AtEveryBibitem{\clearfield{eprint}}
\AtEveryBibitem{\clearfield{note}}

\title{\LARGE \bf

Quasi-Static Fault-Tolerant Feedback Control of a Quadrotor Carrying a Slung Payload under Rotor Failure

}

\author{Mohamed Al Lawati$^{1}$ and Adeel Akhtar$^{2}$%
\thanks{$^{1}$ Department of Mechanical and Industrial Engineering, Sultan Qaboos University, Muscat, Oman. {\tt\small mlawati@squ.edu.om}}
\thanks{$^{2}$ Department of Mechanical and Industrial Engineering, New Jersey Institute of Technology, NJ, USA. {\tt\small adeel.akhtar@njit.edu}}
}

\begin{document}

\maketitle
\thispagestyle{empty}
\pagestyle{empty}

\begin{abstract} 
This paper presents a nonlinear control law for a slung payload system (SPS) consisting of a quadrotor unmanned aerial vehicle (UAV) carrying a slung payload. We propose a fault-tolerant control (FTC) design for the case of a complete single-rotor failure, i.e., only three functioning propellers. The failure is not modeled as a disturbance to be handled by estimation. Rather, we model the faulty SPS in control-affine form and design a quasi-static feedback law directly for the faulty dynamics. The controller achieves local exponential convergence of the payload position to a geometric path in space, rather than tracking a timed reference trajectory. Moreover, the transverse dynamics are exponentially stable, and the path-following manifold is rendered controlled invariant by feedback. Controlled invariance allows motion along the path to be prescribed independently through a tangential command while the state stays on the path-following manifold. The UAV body angular velocity remains bounded under bounded applied torque despite the failure. To validate the control design, numerical simulations apply the controller derived from the faulty SPS model in coordinates to the coordinate-free SPS model. Code and animations are available at \url{https://gradslab.github.io/qs-sl/}.

\end{abstract}

\section{Introduction}
Payload transportation is an important application of multirotor unmanned aerial vehicles (UAVs)~\cite{LvYuZha26}. Conventional gripper-based designs keep the payload close to the UAV, potentially restricting its maneuverability and complicating cooperative transportation by multiple UAVs. Slung payload systems (SPSs) offer an attractive alternative by suspending the payload beneath the UAV~\cite{SreLeeKum13,AkhSalSha21}. However, this configuration introduces strongly coupled vehicle--payload dynamics, pendulum oscillations, and additional underactuation, making feedback control considerably more challenging, particularly under rotor failure. In this paper, we address fault-tolerant path following of an SPS under complete single-rotor failure, with the objective of ensuring payload convergence to a prescribed geometric path while allowing independent control of motion along it.

Motion control for SPSs has been studied in the literature. Early work of \cite{PalFieCru12} plans swing-free trajectories for SPSs, but without addressing a feedback controller. A year later, the authors in~\cite{SreLeeKum13} proved differential flatness of the SPS and designed a coordinate-free feedback law using three nested control loops. However, gain tuning is not intuitive and implementation of the proved nested control structure requires approximation of inter-loop reference signals. Furthermore, only exponential attractiveness is guaranteed as opposed to exponential stability. The subsequent work of \cite{AkhSalSha21, AkhSalSha22} utilizes feedback linearization and provides simpler gain tuning mechanism that leads to exponential stability in a set stabilization setting. However, the SPS model assumed constant tension in the cable equal to tension at hover. This assumption is not realistic especially in fast maneuvers. The series of work in~\cite{LawLyn23,LawJiaLyn23,LawLyn23aim} models the cable tension as a function of state variables, and proposes feedback linearizing controllers in trajectory-tracking~\cite{LawJiaLyn23} and set-stabilization~\cite{LawLyn23} settings. Payload mass uncertainties are addressed in~\cite{LawLyn23aim} numerically, but without compensator design. The work of~\cite{KanYuaSha25} looks at payload mass uncertainty more carefully and proposed an estimator integrated in the nested control loop of~\cite{SreLeeKum13}. The article~\cite{KanShaAlk25} proposes a robust controller that rejects bounded matched disturbances, i.e.,  disturbances entering through the UAV thrust force channel. However, their analysis is limited to set-point regulation and does not account for UAV attitude dynamics. In~\cite{AlkWanSha26}, the authors study SPS trajectory-tracking problem in the case of bounded disturbances and thrust saturation. Using control barrier functions, the work of~\cite{RezKanHar26} addresses set-point regulation capable of avoiding obstacles. However, all works \cite{KanShaAlk25, AlkWanSha26,RezKanHar26} do not account for UAV rotational dynamics in control design. For further work in SPS motion control, the reader is referred to the recent review article~\cite{LvYuZha26}.

\begin{figure}[t]
  \centering
  \begin{minipage}[b]{0.48\linewidth}
    \centering
    \def\svgwidth{\columnwidth}
    \begingroup%
  \makeatletter%
  \providecommand\color[2][]{%
    \errmessage{(Inkscape) Color is used for the text in Inkscape, but the package 'color.sty' is not loaded}%
    \renewcommand\color[2][]{}%
  }%
  \providecommand\transparent[1]{%
    \errmessage{(Inkscape) Transparency is used (non-zero) for the text in Inkscape, but the package 'transparent.sty' is not loaded}%
    \renewcommand\transparent[1]{}%
  }%
  \providecommand\rotatebox[2]{#2}%
  \newcommand*\fsize{\dimexpr\f@size pt\relax}%
  \newcommand*\lineheight[1]{\fontsize{\fsize}{#1\fsize}\selectfont}%
  \ifx\svgwidth\undefined%
    \setlength{\unitlength}{235.19311523bp}%
    \ifx\svgscale\undefined%
      \relax%
    \else%
      \setlength{\unitlength}{\unitlength * \real{\svgscale}}%
    \fi%
  \else%
    \setlength{\unitlength}{\svgwidth}%
  \fi%
  \global\let\svgwidth\undefined%
  \global\let\svgscale\undefined%
  \makeatother%
  \begin{picture}(1,0.77368179)%
    \lineheight{1}%
    \setlength\tabcolsep{0pt}%
    \put(0,0){\includegraphics[width=\unitlength,page=1]{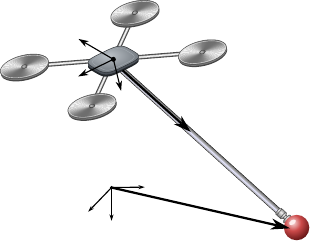}}%
    \put(0.2,0.67){\makebox(0,0)[lt]{\lineheight{1.25}\smash{\begin{tabular}[t]{l}$b_2$\end{tabular}}}}%
    \put(0.19,0.49){\makebox(0,0)[lt]{\lineheight{1.25}\smash{\begin{tabular}[t]{l}$b_1$\end{tabular}}}}%
    \put(0.39,0.43){\makebox(0,0)[lt]{\lineheight{1.25}\smash{\begin{tabular}[t]{l}$b_3$\end{tabular}}}}%
    \put(0.45157666,0.6422865){\makebox(0,0)[lt]{\lineheight{1.25}\smash{\begin{tabular}[t]{l}$m_q, J$\end{tabular}}}}%
    \put(0.645658,0.37440896){\makebox(0,0)[lt]{\lineheight{1.25}\smash{\begin{tabular}[t]{l}$q$\end{tabular}}}}%
    \put(0.72818468,0.27015602){\makebox(0,0)[lt]{\lineheight{1.25}\smash{\begin{tabular}[t]{l}$L$\end{tabular}}}}%
    \put(0.95,0.1){\makebox(0,0)[lt]{\lineheight{1.25}\smash{\begin{tabular}[t]{l}$m_q$\end{tabular}}}}%
    \put(0.47979865,0.16343249){\makebox(0,0)[lt]{\lineheight{1.25}\smash{\begin{tabular}[t]{l}$n_1$\end{tabular}}}}%
    \put(0.19007337,0.05600039){\makebox(0,0)[lt]{\lineheight{1.25}\smash{\begin{tabular}[t]{l}$n_2$\end{tabular}}}}%
    \put(0.38321057,0.06019438){\makebox(0,0)[lt]{\lineheight{1.25}\smash{\begin{tabular}[t]{l}$n_3$\end{tabular}}}}%
    \put(0.7,0.02){\makebox(0,0)[lt]{\lineheight{1.25}\smash{\begin{tabular}[t]{l}$p$\end{tabular}}}}%
  \end{picture}%
\endgroup%
\\
    \footnotesize (a) Healthy rotors.
  \end{minipage}\hfill
  \begin{minipage}[b]{0.48\linewidth}
    \centering
    \def\svgwidth{\columnwidth}
    \begingroup%
  \makeatletter%
  \providecommand\color[2][]{%
    \errmessage{(Inkscape) Color is used for the text in Inkscape, but the package 'color.sty' is not loaded}%
    \renewcommand\color[2][]{}%
  }%
  \providecommand\transparent[1]{%
    \errmessage{(Inkscape) Transparency is used (non-zero) for the text in Inkscape, but the package 'transparent.sty' is not loaded}%
    \renewcommand\transparent[1]{}%
  }%
  \providecommand\rotatebox[2]{#2}%
  \newcommand*\fsize{\dimexpr\f@size pt\relax}%
  \newcommand*\lineheight[1]{\fontsize{\fsize}{#1\fsize}\selectfont}%
  \ifx\svgwidth\undefined%
    \setlength{\unitlength}{235.19311523bp}%
    \ifx\svgscale\undefined%
      \relax%
    \else%
      \setlength{\unitlength}{\unitlength * \real{\svgscale}}%
    \fi%
  \else%
    \setlength{\unitlength}{\svgwidth}%
  \fi%
  \global\let\svgwidth\undefined%
  \global\let\svgscale\undefined%
  \makeatother%
  \begin{picture}(1,0.77368179)%
    \lineheight{1}%
    \setlength\tabcolsep{0pt}%
    \put(0,0){\includegraphics[width=\unitlength,page=1]{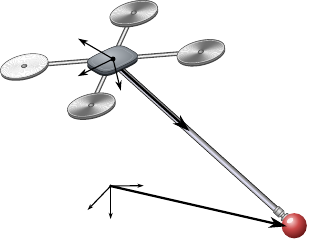}}%
    \put(0.2,0.67){\makebox(0,0)[lt]{\lineheight{1.25}\smash{\begin{tabular}[t]{l}$b_2$\end{tabular}}}}%
    \put(0.19,0.49){\makebox(0,0)[lt]{\lineheight{1.25}\smash{\begin{tabular}[t]{l}$b_1$\end{tabular}}}}%
    \put(0.39,0.43){\makebox(0,0)[lt]{\lineheight{1.25}\smash{\begin{tabular}[t]{l}$b_3$\end{tabular}}}}%
    \put(0.45157666,0.6422865){\makebox(0,0)[lt]{\lineheight{1.25}\smash{\begin{tabular}[t]{l}$m_q, J$\end{tabular}}}}%
    \put(0.645658,0.37440896){\makebox(0,0)[lt]{\lineheight{1.25}\smash{\begin{tabular}[t]{l}$q$\end{tabular}}}}%
    \put(0.72818468,0.27015602){\makebox(0,0)[lt]{\lineheight{1.25}\smash{\begin{tabular}[t]{l}$L$\end{tabular}}}}%
    \put(0.95,0.1){\makebox(0,0)[lt]{\lineheight{1.25}\smash{\begin{tabular}[t]{l}$m_q$\end{tabular}}}}%
    \put(0.47979865,0.16343249){\makebox(0,0)[lt]{\lineheight{1.25}\smash{\begin{tabular}[t]{l}$n_1$\end{tabular}}}}%
    \put(0.19007337,0.05600039){\makebox(0,0)[lt]{\lineheight{1.25}\smash{\begin{tabular}[t]{l}$n_2$\end{tabular}}}}%
    \put(0.38321057,0.06019438){\makebox(0,0)[lt]{\lineheight{1.25}\smash{\begin{tabular}[t]{l}$n_3$\end{tabular}}}}%
    \put(0.7,0.02){\makebox(0,0)[lt]{\lineheight{1.25}\smash{\begin{tabular}[t]{l}$p$\end{tabular}}}}%
  \end{picture}%
\endgroup%
\\
    \footnotesize (b) Single-rotor failure.
  \end{minipage}
  \caption{The slung payload system. %
  }
  \label{fig:SPS}
\end{figure}

For an SPS with a quadrotor UAV, there is no actuator redundancy, so the failure of rotor(s) immediately jeopardizes the whole flight unless the controller has been designed to handle failure situations. Within this stream, the work of~\cite{YuWuHe24} studies SPSs under partial UAV propeller damage, i.e., partially functioning propellers. The propeller damage is viewed as injecting disturbance to the SPS, and thus, a disturbance observer is designed. Likewise, \cite{LiLiYan24} compensates actuator faults of an SPS with a disturbance observer with the help of a neural network. The article~\cite{YiJiaTia26} extends the disturbance observer to the case of two UAVs carrying a common payload. Instead of an observer, \cite{LuoLiZha26} lumps the effect of partial actuator faults into an unknown bounded term whose bound is estimated  by adaptive laws. In our work, however, the failure is not treated as a disturbance or an uncertainty to be estimated and compensated for. Rather, the controller is designed directly for the faulty dynamics, hence, bypassing any fault estimation step.  Furthermore, we study the case of a complete failure of a single rotor as opposed to partial faults.

The main contributions of this paper are summarized as follows:
\begin{enumerate}
\item \textbf{Quasi-static feedback design:} We develop a closed-form fault-tolerant QSF control law~\eqref{eq:ut},~\eqref{eq:tau23} that decouples the transverse and tangential dynamics. Lemma~\ref{lem:Atilde} establishes the invertibility of the associated decoupling matrix on $\mathcal X_d$.

\item \textbf{Exponential stability and controlled invariance:} Theorem~\ref{thm:main_result} establishes exponential stability of the transverse dynamics and controlled invariance of the path-following manifold. The tangential input $\nu_3$ independently assigns motion along the prescribed geometric path.

\item \textbf{Bounded UAV angular velocity:} Lemma~\ref{lma:OM_bound} establishes input-to-state stability of the UAV rotational dynamics. Consequently, Theorem~\ref{thm:main_result} guarantees bounded body angular velocity under bounded applied torque, despite complete single-rotor failure.
\end{enumerate}
Finally, to the best of the authors' knowledge, this is the first work to propose a closed-form fault-tolerant feedback control law for a quadrotor slung payload system under complete single-rotor failure.

The paper is organized as follows. Section~\ref{sec:modl} introduces the model of healthy and faulty SPSs. The problem is stated in Section~\ref{sec:problem}. Section~\ref{sec:fdk} presents the controller design, and stability analysis is presented in Section~\ref{sec:stbility}. Section~\ref{sec:simulations} reports the MATLAB simulations, and Section~\ref{sec:conc} concludes the paper.

\section{Modeling} \label{sec:modl}
The SPS consists of a UAV and a taut cable carrying slung payload. The taut cable is modeled by a massless  pendulum pivoted at the UAV's CoM by a universal joint where the payload is rigidly attached to the other tip of the pendulum. The SPS moves in a north-east-down navigation frame $\cN = \{n_1, n_2,n_3\}$ whose origin is $O_{\cN}$. We define a body-fixed frame  $\cB = \{b_1, b_2, b_3\}$ whose origin $O_{\cB}$ is fixed at the UAV CoM such that $b_1$ points to the UAV heading direction and $b_3$ points down when the UAV is at hover, i.e., horizontally oriented. The heading direction of the UAV is along $b_1$. As shown in Fig.~\ref{fig:SPS}, the setup does not change between the healthy SPS and the SPS under single-rotor failure.

The payload position and velocity relative to $\cN$ are
$p \in \R^3$ and $v = \dot p \in \R^3$, respectively. The attitude of $\cB$ relative to $\cN$ is represented by a rotation matrix $R \in SO(3)$ with columns $b_1$, $b_2$, $b_3$,
and the body-frame angular velocity is $\Omega \in \R^3$. The pendulum attitude is described by a unit vector $q \in \rS^2$  and its angular velocity is denoted by $\omega \in \cW$ where $\cW = \{ \omega \in \R^3: \omega \cdot q =0 \}$. Note that $q$ points from the UAV CoM to the payload. 

The SPS configuration evolves on an 8-dimensional manifold, while the full state $(p,q,R,v,\omega,\Omega) \in \R^3 \times \rS^2 \times SO(3) \times \R^3 \times \cW \times \R^3$ is 16-dimensional. Moreover, it
is actuated by a 4-dimensional input $u_n = (u_t, \tau)$, where $u_t \in (0,\infty)$ is the thrust along $-b_3$ direction and $\tau \in \R^3$ is the UAV body torque with $i^{\text{th}}$ component $\tau_i$ acting about
$b_i$, $i = 1, 2, 3$. The system is thus underactuated with four degrees of underactuation. The UAV mass is $m_q$ and its moment of
inertia is $J$, which is assumed diagonal. The pendulum rod is assumed massless. The payload is a point mass with mass $m_p$. Following~\cite{SreLeeKum13}, with total mass $M = m_p + m_q$ and gravitational acceleration $\g$, the SPS dynamics are
\begin{subequations}\label{eq:SPS_geo}
  \begin{align}
    \dot{p} &= v \label{eq:pos}\\
    \dot{v} &= -\frac{1}{M}\left( m_qL \norm{\omega}^2 + u_t (b_3 \cdot q) \right)q + \g e_3 \label{eq:vel} \\
    \dot q  &= \omega \times q \label{eq:q_dt}\\
    \dot \omega &= \frac{u_t}{m_qL}(q\times b_3) \label{eq:omega_dt}\\
    \dot R &= R \hat \Omega \label{eq:R_dt}\\
    \dot \Omega &= J^{-1}(\tau - \Omega \times J\Omega)  - k_{\Omega}\Omega \label{eq:Om_dt}
  \end{align}
\end{subequations}
where $k_{\Omega} \in \R^{3\times 3}$ is diagonal with positive diagonal entries and $k_{\Omega}\Omega$ is a damping term adapted from~\cite{FaeFraSca17, KaiAllHua17}.

While~\eqref{eq:SPS_geo} provides a coordinate-free description of the SPS
dynamics, control design and simulation are often more convenient in local
coordinates. We parameterize $R \in SO(3)$ using the $2$-$1$-$3$
Euler angle sequence $\eta = (\theta,\phi,\psi)^\top$, defined by
$R = R_y(\theta) R_x(\phi) R_z(\psi)$, and we parameterize $q \in \rS^2$ using
the spherical coordinates $(\alpha,\beta)$. Explicitly,
\begin{subequations}\label{eq:coord_param}
\begin{align}
R(\eta) &=
\begin{bmatrix}
s_\theta s_\phi s_\psi + c_\theta c_\psi & s_\theta s_\phi c_\psi - c_\theta s_\psi & s_\theta c_\phi\\
c_\phi s_\psi & c_\phi c_\psi & -s_\phi\\
c_\theta s_\phi s_\psi - s_\theta c_\psi & c_\theta s_\phi c_\psi + s_\theta s_\psi & c_\theta c_\phi
\end{bmatrix}, \label{eq:R213}\\[4pt]
q(\alpha,\beta) &=
\begin{bmatrix} s_\beta \\ -s_\alpha c_\beta \\ c_\alpha c_\beta \end{bmatrix}, \label{eq:q}
\end{align}
\end{subequations}
where $s_x := \sin x$, $c_x := \cos x$, and $t_x := \tan x$ throughout. From~\eqref{eq:R213}, the UAV body-frame $b_3$-axis expressed in $\cN$ is
\begin{equation}\label{eq:b3}
b_3(\eta) = \begin{bmatrix} s_\theta c_\phi & -s_\phi & c_\theta c_\phi \end{bmatrix}^\top.
\end{equation}
 
Since $\omega \in \cW$ and $\norm{q} = 1$, we may recover $\omega$ from $\dot q$ via
$\omega = q \times \dot q$. Let $\gamma_\alpha:=\dot \alpha$ and  $\gamma_\beta:=\dot \beta$ represent the Euler rates of the pendulum. Let $\gamma := [\gamma_\alpha,\gamma_\beta]^\top$. Then,
\begin{equation}
    \omega = W_\gamma(\alpha,\beta) \gamma \quad \text{where} \quad
     W_\gamma = \begin{bmatrix}
         c_\beta^2 & 0 \\
         s_\alpha s_{2\beta}/2 & c_\alpha \\
        -c_\alpha s_{2\beta}/2 & s_\alpha
     \end{bmatrix}
\end{equation}
Away from $\cos\beta=0$, $W_\gamma$ is full column rank and has a left pseudoinverse $W_\gamma^+ = \left(W_\gamma^\top W_\gamma\right)^{-1}W_\gamma^\top$. Hence, $\gamma = W_\gamma^+ \omega$ yielding $\dot \gamma = \dot W_\gamma^+ \omega + W_\gamma^+ \dot\omega$. Using~\eqref{eq:omega_dt}, we get
\begin{equation} \label{eq:gamma_dt}
    \dot \gamma = 
    \begin{bmatrix}
        2\gamma_\alpha\gamma_\beta t_\beta 
        + \frac{(q \times b_3)^\top e_1}{m_qLc_\beta^2} u_t \\
        -s_\beta c_\beta \gamma_\alpha^2 
        - \frac{\left(q\times(q \times b_3)\right)^\top e_1}{m_qLc_\beta} u_t
    \end{bmatrix}.
\end{equation}

Turning to the UAV, the kinematic relation between the Euler rates, $\dot\eta$, and body rate, $\Omega$, is standard~\cite{SpoHutVid20}. For 2-1-3 Euler, we have
\begin{equation}\label{eq:eta_dt}
  \dot \eta = W\Omega  \quad \text{where} \quad W = \begin{bmatrix}
\dfrac{s_\psi}{c_\phi} & \dfrac{c_\psi}{c_\phi} & 0\\[6pt]
c_\psi & -s_\psi & 0\\[3pt]
t_\phi\, s_\psi & t_\phi\, c_\psi & 1
\end{bmatrix}.
\end{equation}

We are now ready to write~\eqref{eq:SPS_geo} in coordinates.
Let the state vector be $
x = \big(p, \alpha, \beta, \eta, v, \gamma,
\Omega\big) \in \R^{16}$, and let input vector be
 $u_n = (u_t,\tau) \in \R^4$. The dynamic equation \eqref{eq:omega_dt} is replaced by~\eqref{eq:gamma_dt}, and the kinematics~\eqref{eq:R_dt} is replaced by~\eqref{eq:eta_dt}. The kinematics~\eqref{eq:q_dt} is simply replaced by $\dot{\alpha}=\gamma_\alpha$ and $\dot{\beta}=\gamma_\beta$. Using the fact that $q$, $b_3$ are written in terms of $x$ using \eqref{eq:q}, \eqref{eq:b3}, and $\norm{\omega}^2 = \gamma_\alpha^2 c_\beta^2 + \gamma_\beta^2$, we obtain

\begin{equation}\label{eq:SPS_healthy}
\dot x = f(x) + G_n(x)u_n,
\end{equation}
with
\begin{align}
f(x) \!&=\!
\begin{bmatrix}
v\\
\gamma\\
W\,\Omega\\
-\dfrac{m_qL}{M}\norm{\omega}^2\,
  q + \g e_3\\
2\gamma_\alpha\gamma_\beta t_\beta\\
-\gamma_\alpha^2 s_\beta c_\beta\\
-J^{-1}\big(\Omega \times J\Omega\big) - k_{\Omega}\Omega
\end{bmatrix}\!\!,
G_n(x) \!=\!
\begin{bmatrix}
0_{8\times1} & 0_{8\times3}\\
g_t(x) & 0_{5\times3}\\
0_{3\times1} & J^{-1}
\end{bmatrix}\!\!,
\end{align}
where
\begin{equation}
g_t(x) =
\begin{bmatrix}
-\dfrac{1}{M}\left(b_3^\top q\right)q\\[4pt]
\dfrac{(q \times b_3)^\top e_1}{m_qLc_\beta^2}\\[4pt]
- \dfrac{(q\times(q \times b_3))^\top e_1}{m_qLc_\beta} 
\end{bmatrix} \in \R^5.
\end{equation}
The tuple $(\alpha,\beta,\gamma) \in \R^4$
provides a minimal, unconstrained representation of the pendulum
configuration and velocity, in contrast to $(q,\omega) \in \rS^2 \times \cW$, which is subject to the algebraic constraints $\|q\| = 1$ and $\omega^\top q = 0$. Similarly, $\eta \in \R^3$ is a coordinate representation of $R$, which must adhere to the constraints $R^\top R=RR^\top=I_3$, $\det (R)=1$. Consequently, $x \in \R^{16}$ in~\eqref{eq:SPS_healthy} is a chart-based, minimal-dimensional realization of the $16$-dimensional
state space $\R^3 \times \rS^2 \times SO(3) \times \R^3 \times \cW \times \R^3$, valid away from the coordinate singularities $\phi = \pm\pi/2$ and $\beta = \pm\pi/2$.

Let $f_i \geq 0$ denote the thrust generated by rotor $i$, $i=1,2,3,4$. Let $f:=\col{f_1,f_2,f_3,f_4}$. Following~\cite{MahAliAkh2025}, the relation between $f$ and $u_n = \col{u_t,\tau}$ is $u_n = Q f$, where 
\begin{equation*}
    Q = 
    \begin{bmatrix}
    1 & 1 & 1 & 1\\
    0 & -\ell & 0 & \ell\\
    -\ell & 0 & \ell &0\\
    c_d & -c_d & c_d& -c_d
    \end{bmatrix},
\end{equation*}
where $\ell$ is the rotor-to-CoM distance of the UAV, and $c_d$ is the propellers drag-to-thrust ratio. Without loss of generality, let rotor 2 be completely failed, i.e., $f_2\equiv0$.
For fixed total thrust, single-rotor failure restricts $\tau$ to an affine plane in $\R^3$. We can write $\tau_{1} = c_1 u_t - c_2 \tau_{3}$, where $c_1 = \ell/2$, $c_2 = \ell/(2c_d)$. The resulting input vector reduces to $u = (u_t,\tau_2,\tau_3)$. Hence, system~\eqref{eq:SPS_healthy} turns into
\begin{equation}\label{eq:SPS_faulty}
    \dot x = f(x) + G(x)u
\end{equation}
where $f(x)$ directly comes from~\eqref{eq:SPS_healthy} and 
\begin{equation*}
    G(x) = 
    \left[
        \begin{array}{cc}
        \multicolumn{2}{c}{0_{8\times 3}} \\[1pt]
        \hdashline\\[-10pt]
        g_t(x) & 0_{5\times 2} \\[1pt]
        \hdashline\\[-10pt]
        \multicolumn{2}{c}{J^{-1}C}
        \end{array}
        \right]\!\!, \text{and }  C =  \begin{bmatrix}
            c_1 & 0 & -c_2 \\
            0   & 1 & 0    \\
            0   & 0 & 1
        \end{bmatrix}.
\end{equation*}
For the remaining rotors to generate nonnegative thrust, the commanded inputs must satisfy $u_t-\tau_3/c_d\ge0$ and $u_t+\tau_3/c_d\ge 2|\tau_2|/\ell$. Their satisfaction must be checked separately from the feedback-linearization analysis. Subsequent control derivation is based on model~\eqref{eq:SPS_faulty}. However, numerical simulations discussed in Section~\ref{sec:simulations} are performed using the dynamics~\eqref{eq:SPS_geo}.

\section{Problem Formulation}\label{sec:problem}

The primary goal is to steer the payload position onto a geometric path $\mathcal{C} \subset \mathbb{R}^3$ and maintain payload's motion invariance along $\cC$. A secondary objective is to control payload's motion along $\cC$. These two goals are to be achieved despite a complete single-rotor failure. By a geometric curve, we mean a spatial curve that is not parameterized by time. By invariance we mean that when the complete initial state belongs to the path-following manifold, the payload remains on the path. In particular, the controller eliminates off-path transients for states initialized on this manifold. This property holds despite a complete single-rotor failure. Conventional trajectory-tracking controllers cannot offer this invariance guarantee. 

We describe the curve $\mathcal{C}\subset\mathbb{R}^3$ as the intersection of two independent surfaces:
\begin{equation}\label{eq:C}
\mathcal{C} = \{p\in\mathbb{R}^3: h_1(p)=h_2(p)=0,\; \d h_1(p)\times \d h_2(p)\neq 0\}.
\end{equation}
Following~\cite{AkhWasNie2013}, we do not require $\mathcal{C}$ to be closed, which relaxes the closedness assumption made in related work~\cite{RozMag2012}. The three-dimensional input $u$ has enough control authority to enforce the constraints $h_1(p)=h_2(p)=0$, which we call the transverse constraints. The remaining degree of actuation in $u$ is used to specify motion along $\cC$. We introduce a third output $h_3(p)$ as a smooth local path-coordinate function, for example the arc length along $\mathcal{C}$ measured from a reference point. In contrast to $h_1$ and $h_2$, the output $h_3$ is generally not driven to zero; it is instead used to prescribe the desired motion along $\mathcal{C}$. To simplify subsequent calculations, we consider curves with constant height, $r^\ast$. The analysis uses a regular local chart; a closed curve can be covered by multiple such charts.
Thus, the output map is
\begin{equation}\label{eq:output}
h(p)=\mathrm{col}\bigl(h_1(p_1,p_2),p_3 - r^\ast,h_3(p)\bigr).
\end{equation}
Since~\eqref{eq:C} requires $\nabla h_1\times\nabla h_2\neq 0$ on
$\mathcal C$, the vector $\nabla h_1\times\nabla h_2$ is tangent to
$\mathcal C$. Because $h_3$ is a path coordinate, its derivative along
this tangent is nonzero, that is,
$\langle\nabla h_1\times\nabla h_2,\nabla h_3\rangle\neq 0$ on
$\mathcal C$. This condition is open, so by continuity there exists an
open neighborhood of $\mathcal C$ on which it holds. We fix such a
neighborhood and denote it by
\begin{equation*}
\mathcal N_{\mathcal C}:=\big\{p\in\mathbb R^3:\
\langle\nabla h_1\times\nabla h_2,\nabla h_3\rangle\neq 0\big\}.
\end{equation*}
We consider only payload positions $p\in\mathcal N_{\mathcal C}$, where
the outputs~\eqref{eq:output} are regular. We assume that the states evolve in the domain $$\mathcal X:=\{x\in\mathbb R^{16}:\ p\in\mathcal N_{\mathcal C},\ \alpha,\beta, \phi \in(-\tfrac{\pi}{2},\tfrac{\pi}{2}),\ \langle b_3,q\rangle\neq0\}.$$ The condition $\langle b_3,q\rangle\neq 0$ excludes
configurations in which the pendulum  is orthogonal to the thrust
axis, where the thrust has no authority over the payload acceleration. When  $\langle b_3,q\rangle > 0 $ or $\langle b_3,q\rangle < 0 $, the pendulum respectively points below or above UAV's ($b_1$-$b_2$)-plane. With the above setup, we are ready to state our problem. %

\begin{problem}[Fault-tolerant Path Following Problem (Ft-PFP)]
\label{prob} 
Consider an SPS with a completely failed rotor~\eqref{eq:SPS_faulty} and a path~\eqref{eq:C}. Design a feedback controller $\kappa: \mathcal X_d \to \R^3$, $x\mapsto \kappa(x) = u$, so that the closed-loop SPS meets the following goals:
\begin{enumerate}[label=\textbf{G\arabic*} ]
	\item \label{goal:1} The payload converges to $\cC$ exponentially.
	\item \label{goal:2} Payload's motion along $\cC$ is controlled-invariant.
    \item \label{goal:3} The payload follows a prescribed tangential motion along $\mathcal C$.
    \item \label{goal:4} The UAV body angular velocity stays bounded under bounded applied torque.
\end{enumerate}
\end{problem}
We remark that fault detection and isolation (FDI) is beyond the scope of this work, and a recent article proposing an FDI for an SPS is~\cite{ZhuLiYin24}. We proceed with a solution to Problem~\ref{prob} in the following section.

\section{Solution using a QSF}\label{sec:fdk}
Our approach splits the output coordinates into transverse and tangential parts. The transverse outputs $h_1$ and $h_2$ describe the desired path $\mathcal{C}$ and are used to ensure both convergence to the path and its invariance. The third output $h_3=s(p)$ parametrizes motion along $\mathcal{C}$ and, in general, is not driven to zero. It is instead used to specify the desired tangential motion along the path, for instance point stabilization, velocity tracking, or acceleration tracking. As a result, the origin of transverse dynamics is stabilized while the tangential dynamics remain freely assignable through a user-specified tangential  command.

Consider system~\eqref{eq:SPS_faulty} and output map~\eqref{eq:output}. Let's differentiate~\eqref{eq:output} with respect to time until an input appears. This process gives
\begin{equation*}
    \col{\ddot h_1, \ddot h_2,\ddot h_3} = Au  +  \col{L_f^2h_1,L_f^2h_2,L_f^2h_3}
\end{equation*}
where $A \in \R^{3\times 3}$ is the decoupling matrix whose $i^\text{th}$ row is $$-\frac{\dotp{b_3}{q}}{M}\, [\dotp{q}{\nabla h_i}, \, 0,\, 0].$$ The matrix $A$ has rank one when $\langle b_3,q\rangle\neq0$ and at least one $\langle q,\nabla h_i\rangle$ is nonzero. Both conditions hold on $\cX$, since $\langle q,\nabla h_2\rangle=c_\alpha c_\beta>0$. Hence, we can establish a single independent equation between $\ddot h$ and $u$. Without loss of generality, we relate between $\ddot h_2$ directly to $u$. In particular, we have $\ddot h_2 = -\frac{\dotp{b_3}{q}}{M}\, \dotp{q}{\nabla h_2}u_t + L_f^2h_2$. Since $u_t$ can be used to directly affect $\ddot h_2$, we can easily stabilize the $h_2$-dynamics by introducing an input transformation. Let $\nu_2 := \ddot h_2$ be a transformed input. Then, we have
\begin{equation}\label{eq:nu_2}
    \nu_2 = \ddot h_2 = -\frac{\dotp{b_3}{q}}{M}\, \dotp{q}{\nabla h_2}u_t + L_f^2h_2.
\end{equation}
To design $\nu_2$, we realize that $\ddot h_2=\nu_2$ is a simple double integrator. We define $T_2(x) := \col{h_2,\dot h_2}$ and $z^\pitchfork_2 := T_2(x)$. Thus, we can rewrite  $\ddot h_2=\nu_2$ as
\begin{equation}\label{eq:ztransverse_2}
    \dot z^\pitchfork_2 = A_{c2}z^\pitchfork_2  + B_{c2}\nu_2
\end{equation}
where $(A_{c2}, B_{c2})$ is a 2-by-2 Brunovsky pair~\cite{MarTom98}. We choose 
\begin{equation}
    \nu_2 = -K_2 z^\pitchfork_2
\end{equation}
where $K_2 \in \R^{1\times 2}$ is a gain matrix chosen such that $A_{c2}-B_{c2}K_2$ is Hurwitz, thus, resulting in exponential stability of the $z_2^\pitchfork$-dynamics. Given the design of $\nu_2$, the thrust can immediately be recovered from~\eqref{eq:nu_2} as
\begin{equation}\label{eq:ut}
     u_t = M\frac{L_f^2h_2 - \nu_2}{\dotp{b_3}{q}c_\alpha  c_\beta}
\end{equation}
where $L_f^2h_2 = -(Lm_q/M)\norm{\omega}^2\c_\alpha\c_\beta + \g$. Hence, $u_t$ is well defined on $\cX$. On $\cX$, the thrust equation~\eqref{eq:ut} mandates $c_\alpha c_\beta \neq 0$, hence $c_\alpha c_\beta$ cannot change sign. In our chosen chart $c_\alpha c_\beta>0$, so the payload lies below the horizontal $(n_1$-$n_2)$-plane.

\begin{remark}\label{rmk:u_t_not_zero}
    Restricting $\langle b_3,q\rangle > 0 $ and $c_\alpha c_\beta >0$ means pendulum's operation mode is below $(b_1$-$b_2)$-plane and below the horizontal plane. For this operation mode, we require  $L_f^2h_2 - \nu_2 > 0$ to ensure $u_t >0$. 
    For subsequent discussion, we will work with $\mathcal X_d:=\{x\in\mathcal X:\ \langle b_3,q\rangle>0,\c_\alpha c_\beta >0,\
L_f^2h_2-\nu_2>0\}$.
\end{remark}

Now, we need to get another two equations for the remaining inputs, namely, $\tau_2$ and $\tau_3$. We take further time derivatives of $\ddot h_1$ and $\ddot h_3$ until we get two independent equations in terms of $\tau_2$ and $\tau_3$. It turns out we need six time derivatives of $h_1$ and $h_3$. After substituting the thrust law~\eqref{eq:ut}, let $G=[g_1\;g_2\;g_3]$ and define $\bar f=f+g_1u_t(x)$, with $\nu_2=-K_2z_2^\pitchfork$, so the drift includes the torque $c_1u_t$ induced by rotor failure. Thus, we compute
\begin{equation}   \label{eq:h1h3_d6t} 
       \begin{bmatrix} h_1^{(6)} \\ h_3^{(6)}\end{bmatrix} =  
        \tilde{A}
        \begin{bmatrix}
            \tau_2 \\ \tau_3
        \end{bmatrix}  
        +
        \begin{bmatrix}
            L_{\bar f}^{6}h_1 \\
            L_{\bar f}^{6}h_3
        \end{bmatrix},
\end{equation}
where
\begin{equation}\label{eq:Atilde}
    \begin{aligned}
    \tilde A &= \sigma
        \begin{bmatrix}
            (\nabla h_1 \times \nabla h_2)^\top \\
            (\nabla h_3 \times \nabla h_2)^\top
        \end{bmatrix}
        \begin{bmatrix}
            \frac{(b_2 \times q) \times q}{J_2} & -c_2\frac{(b_1 \times q) \times q}{J_1}
        \end{bmatrix}
    \\
    \sigma &= -\frac{u_t \, (\g-\nu_2)}{m_q L \, \langle b_3, q \rangle \, (c_\alpha c_\beta)^2}
    \end{aligned}
\end{equation}
To compute $\det \tilde A$, we use the Cauchy--Binet (Lagrange) identity. Since  $\tilde A$ in~\eqref{eq:Atilde} is $2\times 2$, we get $
\det\tilde A=\sigma^2\Big\langle(\nabla h_1\times\nabla h_2)\times(\nabla h_3\times\nabla h_2),\ \tfrac{(b_2\times q)\times q}{J_2}\times\Big(-c_2\tfrac{(b_1\times q)\times q}{J_1}\Big)\Big\rangle$. By the vector triple product identity, we get $(\nabla h_1\times\nabla h_2)\times(\nabla h_3\times\nabla h_2)=-\langle\nabla h_1\times\nabla h_2,\nabla h_3\rangle\,\nabla h_2$. Now, for any $w\in\mathbb R^3$, $(w\times q)\times q=-w+q\langle q,w\rangle$ is minus the component of $w$ orthogonal to $q$. The cross product of two such vectors is therefore parallel to $q$, with $q$-component $\langle q,b_2\times b_1\rangle=-\langle b_3,q\rangle$, so $ \big((b_2\times q)\times q\big)\times\big((b_1\times q)\times q\big)=-\langle b_3,q\rangle\,q$. Taking the scalar product, and substituting $\sigma^2$ from~\eqref{eq:Atilde}, yields
\begin{equation}\label{eq:detAtilde}
\det\tilde A=-\frac{c_2\,u_t^2\,(\g-\nu_2)^2\,\langle\nabla h_1\times\nabla h_2,\nabla h_3\rangle}{J_1J_2\,(Lm_q)^2\,\langle b_3,q\rangle\,(c_\alpha c_\beta)^{3}}.
\end{equation}

\begin{lemma}\label{lem:Atilde}
The matrix $\tilde A$ given
in~\eqref{eq:Atilde} is well defined and invertible on $\mathcal X_d$.
\end{lemma}

\begin{proof}
On $\mathcal X_d$, $\langle b_3,q\rangle >0$ and $c_\alpha c_\beta >0$, so
$\tilde A$ in~\eqref{eq:Atilde} and $\det\tilde A$
in~\eqref{eq:detAtilde} are well defined. Since $c_2,J_1,J_2,L,m_q>0$,
\eqref{eq:detAtilde} shows that $\det\tilde A=0$ only if $u_t=0$,
$\g-\nu_2=0$, or
$\langle\nabla h_1\times\nabla h_2,\nabla h_3\rangle=0$. We rule out each
case. First, by~\eqref{eq:ut}, $u_t=M\,\frac{L_f^2h_2-\nu_2}{\langle b_3,q\rangle\,c_\alpha c_\beta}>0$. Second, since $L_f^2h_2=\g-\frac{Lm_q}{M}\|\omega\|^2c_\alpha c_\beta$, we have $\g-\nu_2=\big(L_f^2h_2-\nu_2\big)+\frac{Lm_q}{M}\|\omega\|^2c_\alpha c_\beta>0$ for every $\omega$, because the first term is positive and the second
is nonnegative. Third,
$\langle\nabla h_1\times\nabla h_2,\nabla h_3\rangle\neq0$ on
$\mathcal N_{\mathcal C}$ by the definition of $\mathcal N_{\mathcal C}$.
Hence, $\det\tilde A\neq0$.
\end{proof}
It can be shown that tension in the cable is $m_p(\g - \nu_2)/(c_\alpha c_\beta)$. Hence, for cable motion below the horizontal plane ($c_\alpha c_\beta>0$), the requirement $\g - \nu_2 > 0$ corresponds to a taut cable.
\begin{remark}
    Solving a similar problem for a UAV without slung payload results in a similar matrix size for $\tilde A$, i.e., 2-by-2. See~\cite{LawAkh26b}.
\end{remark}

Invertibility of $\tilde A$  allows us to introduce the following input transformation
\begin{equation}   \label{eq:nu1nu3} 
       \begin{bmatrix}
           \nu_1 \\ \nu_3
       \end{bmatrix} :=
       \begin{bmatrix} h_1^{(6)} \\ h_3^{(6)}\end{bmatrix} = 
        \tilde{A}
        \begin{bmatrix}
            \tau_2 \\ \tau_3
        \end{bmatrix}  
        +
        \begin{bmatrix}
            L_{\bar f}^{6}h_1 \\
            L_{\bar f}^{6}h_3
        \end{bmatrix},
\end{equation}
where $\nu_1$, $\nu_3$ are transformed inputs. To facilitate the design of $\nu_1$, let's define $T_1(x) := \col{h_1(x), \dot h_1(x), \ddot h_1(x), h_1^{(3)}(x), h_1^{(4)}(x), h_1^{(5)}(x)}$. Let $z_1^\pitchfork := T_1(x)$. Then,
\begin{equation}\label{eq:ztransverse_1}
    \dot z_1^\pitchfork = A_{c6}z_1^\pitchfork + B_{c6}\nu_1,
\end{equation}
where $(A_{c6}, B_{c6})$ is a 6-by-6 Brunovsky pair.
The input $\nu_1$ is designed to stabilize the origin of the $z_1^\pitchfork$-dynamics. One design is 
\begin{equation}
    \nu_1 = -K_1z_1^\pitchfork,
\end{equation}
where $K_1 \in \R^{1\times 6}$ is chosen such that $A_{c6} - B_{c6}K_1$ is Hurwitz. Now, we define the transverse state as $z^\pitchfork := \col{z_1^\pitchfork,z_2^\pitchfork}$. Hence, letting $\nu^\pitchfork := \col{\nu_1, \nu_2}$, the entire transversal dynamics are given by
\begin{equation}\label{eq:zpitchfork}
    \dot  z^\pitchfork = A^\pitchfork z^\pitchfork + B^\pitchfork \nu^\pitchfork
\end{equation}
where $A^\pitchfork$ and $B^\pitchfork$ are block diagonal matrices with blocks $A_{c6}$, $A_{c2}$ and $B_{c6}$, $B_{c2}$; respectively.

For the output $h_3$, we define $T^\parallel(x):= \col{h_3(x), \dot h_3(x), \ddot h_3(x), h_3^{(3)}(x), h_3^{(4)}(x), h_3^{(5)}(x)}$ and $z^\parallel := T^\parallel(x)$. This gives the tangential dynamics
\begin{equation}\label{eq:ztangential}
    \dot z^\parallel = A_{c6}z^\parallel + B_{c6}\nu_3.
\end{equation}
\begin{remark}\label{rmk:tangential_control}
Unlike $z^\pitchfork$-dynamics, the origin of the $z^\parallel$-dynamics is not necessarily stabilized. Instead, $\nu_3$ is used to assign the desired behavior of $z^{\parallel}$ along the path $\mathcal{C}$. For example, to track a desired path speed $v_d(t)$, define $ e_v:=\mathrm{col}\bigl(\dot h_3-v_d,\ddot h_3-\dot v_d,h_3^{(3)}-\ddot v_d,h_3^{(4)}- v_d^{(3)},h_3^{(5)}- v_d^{(4)}\bigr)$, and choose $\nu_3=v_d^{(5)}-K_3e_v$ with a stabilizing $K_3$. Point stabilization or path-coordinate trajectory tracking are obtained similarly by defining the corresponding error in $h_3$ and its derivatives.
\end{remark}

To this end, we can now easily compute the input torques $\tau_2$, $\tau_3$ from~\eqref{eq:nu1nu3} as
\begin{equation}\label{eq:tau23}
       \begin{bmatrix}
            \tau_2 \\ \tau_3
        \end{bmatrix}  
         = 
         \tilde{A}^{-1}\left(\begin{bmatrix}
           \nu_1 \\ \nu_3
       \end{bmatrix} - 
        \begin{bmatrix}
            L_{\bar f}^{6}h_1 \\
            L_{\bar f}^{6}h_3
        \end{bmatrix} \right).
\end{equation}
As a result, the complete QSF for the SPS is given by~\eqref{eq:ut}  and~\eqref{eq:tau23}.

\section{Stability analysis}\label{sec:stbility}

 Let $T^\pitchfork(x) := \col{T_1(x),T_2(x)}$. Define the map $T: \cX_d \to \R^{14} \times \rS^1 \times\R$, $x\mapsto (z^\pitchfork,z^\parallel, \xi)$, where $\xi \in \rS^1 \times\R$ is chosen so that the map $T$ defines a candidate local coordinate transformation. It suffices to pick $\xi$ such that the Jacobian of $T$, denoted $\mathrm{d} T$, is non-singular in some neighborhood.  We choose
\begin{equation}\label{eq:xi}
    \xi =
    \col{\psi,\,J_1\Omega_1+J_3c_2\Omega_3}
    \in \mathbb{S}^1\times\R .
\end{equation}
The mapping $T$ gives local coordinates at any point where $\mathrm dT$ is nonsingular, as guaranteed by the inverse function theorem. We restrict the coordinate-based interpretation to such regular neighborhoods. In the $(z^\pitchfork, z^\parallel, \xi)$-coordinates, the internal dynamics of the closed-loop system is given by
\begin{align}\label{eq:intr_dyn}
\dot{\xi} = 
    \left.\begin{bmatrix}
         e_3^\top W(\eta)\Omega\\
        -[1, 0, c_2] (\Omega \times J\Omega  +  k_\Omega J\Omega ) + c_1 u_t 
    \end{bmatrix}\right|_{x=T^{-1}(z^\pitchfork,z^\parallel,\xi)}.
\end{align}
Note that although $T^{-1}$ might be difficult to compute, it is guaranteed to locally exist since $T$ is a local diffeomorphism. 

In the $(z^\pitchfork, z^\parallel, \xi)$-coordinates, the closed-loop system stabilizes the origin of the $z^\pitchfork$-dynamics. In other words, the closed-loop system stabilizes the  \textit{path-following manifold}
\begin{equation}\label{eq:x-star}
    \cX^\ast := \{x\in\cX_d: T^\pitchfork(x)=0\},
\end{equation}
which specifies the full transverse constraints for SPS motion along $\cC$. Specific motions are determined by the tangential state $z^\parallel$, which is directly controlled by the closed-loop system. However, the $\xi$-dynamics are uncontrolled. To assess the internal motion, a boundedness estimate is useful. To do so, we first need the following Lemma.
\begin{lemma} \label{lma:OM_bound}
    Let 
    \begin{equation} \label{eq:rot_dyn}
        \dot{\Omega} = J^{-1}(\tau - \Omega \times J\Omega) - k_\Omega \Omega,
    \end{equation}
    where $\Omega \in \R^3$ is the state and $\tau \in \R^3$ is the input. Assume $J$ and $k_\Omega$ are diagonal with strictly positive diagonal entries. Then, the dynamics~\eqref{eq:rot_dyn} is input-to-state stable.
\end{lemma}
\begin{proof}
Let $V=\tfrac12\Omega^\top J\Omega$. Since $J=J^\top$,
\begin{align*}
\dot V &= \Omega^\top J\dot\Omega = \Omega^\top\big(\tau-\Omega\times J\Omega-Jk_\Omega\Omega\big) \\
&= \Omega^\top\tau-\Omega^\top(\Omega\times J\Omega)-\Omega^\top Jk_\Omega\Omega = \Omega^\top\tau-\Omega^\top Jk_\Omega\Omega ,
\end{align*}
where the last step uses $\Omega^\top(\Omega\times J\Omega)=0$, because
$\Omega\times J\Omega$ is orthogonal to $\Omega$. Since $J$ and $k_\Omega$
are diagonal, $\Omega^\top Jk_\Omega\Omega=\sum_{i=1}^3J_ik_i\Omega_i^2\ge a\|\Omega\|^2$,
with $a=\min_iJ_ik_i>0$, so $-\Omega^\top Jk_\Omega\Omega\le-a\|\Omega\|^2$. 
By the Cauchy--Schwarz inequality, $\Omega^\top\tau\le\|\Omega\|\,\|\tau\|$. Hence, $\dot V \leq \norm{\Omega}\norm{\tau} - a \norm{\Omega}^2$.
Note that
$(\sqrt a\,\|\Omega\|-\tfrac1{\sqrt a}\,\|\tau\|)^2 \ge 0$ implying $\|\Omega\|\,\|\tau\|\le\tfrac a2\|\Omega\|^2+\tfrac1{2a}\|\tau\|^2$.
Hence, $\dot V \le -\tfrac a2\|\Omega\|^2+\tfrac1{2a}\|\tau\|^2$. Now, let $J_{\min}=\min_iJ_i$ and $J_{\max}=\max_iJ_i$. Because
$V\le\tfrac12J_{\max}\|\Omega\|^2$, we obtain $
\dot V\le-\mu V+\tfrac1{2a}\|\tau\|^2$, where $\mu=\frac{a}{J_{\max}}$, and the comparison lemma~\cite[Lem. 3.4]{Kha02} together with $\int_0^t e^{-\mu(t-s)}ds \leq 1/\mu$,  we arrive at
\begin{equation}\label{eq:uppe_bound_V}
V(t)\le e^{-\mu t}V(0)+\frac{1}{2a\mu}\sup_{0\le r\le t}\|\tau(r)\|^2 .
\end{equation}
Since $\tfrac12J_{\min}\|\Omega(t)\|^2\le V(t)\le\tfrac12J_{\max}\|\Omega(t)\|^2$, using~\eqref{eq:uppe_bound_V}, we get 
$\tfrac12J_{\min}\|\Omega(t)\|^2 \leq e^{-\mu t}V(0)+\frac{1}{2a\mu}\sup_{0\le r\le t}\|\tau(r)\|^2$. Also, $V(t)\le\tfrac12J_{\max}\|\Omega(t)\|^2$ implies $V(0)\le\tfrac12J_{\max}\|\Omega(0)\|^2$. Therefore,  $$\tfrac12J_{\min}\|\Omega(t)\|^2 \leq \tfrac12J_{\max}\|\Omega(0)\|^2 e^{-\mu t}     +\frac{1}{2a\mu}\sup_{0\le r\le t}\|\tau(r)\|^2,$$ which gives 
\begin{align*}
\|\Omega(t)\|^2 & \le \frac{J_{\max}}{J_{\min}}\, e^{-\mu t}\|\Omega(0)\|^2 + \frac{J_{\max}}{a^2 J_{\min}} \sup_{0\le r\le t}\|\tau(r)\|^2\\
\|\Omega(t)\|&\le\sqrt{\tfrac{J_{\max}}{J_{\min}}}\,e^{-\mu t/2}\|\Omega(0)\|
+\frac1a\sqrt{\tfrac{J_{\max}}{J_{\min}}}\,\sup_{0\le r\le t}\|\tau(r)\|.
\end{align*}
Therefore, the system~\eqref{eq:rot_dyn} is input-to-state stable~\cite[Def. 4.7]{Kha02}.
With $\tau\equiv0$, we get $\|\Omega(t)\|\le\sqrt{\tfrac{J_{\max}}{J_{\min}}}\,e^{-\mu t/2}\|\Omega(0)\|$ and, hence, the origin becomes globally exponentially stable.
\end{proof}
Note that Lemma~\ref{lma:OM_bound} shows that the damping term $k_\Omega\Omega$ is what keeps $\Omega$ bounded.  Without it, $\dot V=\Omega^\top\tau$ has no dissipation, which matters since the rotor failure injects the persistent term $c_1u_t$ into the UAV heading channel. 

Under the QSF law \eqref{eq:ut} and \eqref{eq:tau23}, the transverse dynamics reduce to \eqref{eq:zpitchfork}, whose matrix is Hurwitz by proper gain selection, so $z^{\pitchfork}$ decays exponentially regardless of $z^{\parallel}$ and $\xi$. The tangential state $z^{\parallel}$ is assigned by $\nu_3$. The stability of the overall system cannot be immediately concluded even though $z^{\pitchfork}$ decays exponentially, since the internal dynamics of $\xi$ are uncontrolled. These dynamics must be analyzed, and the next theorem provides the resulting guarantees.

\begin{theorem}\label{thm:main_result}
Consider~\eqref{eq:SPS_faulty} under the QSF law~\eqref{eq:ut},~\eqref{eq:tau23}. Let $K_1,K_2$ make $A_{c6}-B_{c6}K_1$ and $A_{c2}-B_{c2}K_2$ Hurwitz, and let $\nu_3$ be locally Lipschitz. For any closed-loop solution evolving in $\mathcal X_d$, the transverse dynamics are exponentially stable and the path-following manifold $\mathcal X^*$ is controlled invariant. Moreover, if the applied torque is bounded, then the UAV body angular velocity is bounded.
\end{theorem}

\begin{proof}
By the design in Section~\ref{sec:fdk}, $\dot z^{\pitchfork}=A_{\rm cl}z^{\pitchfork}$ with $A_{\rm cl}:=\mathrm{diag}(A_{c6}-B_{c6}K_1,\,A_{c2}-B_{c2}K_2)$, which is Hurwitz. Hence $z^{\pitchfork}=0$ is exponentially stable. If $x(0)\in\mathcal X^*$, then $z^{\pitchfork}(0)=0$, so $z^{\pitchfork}(t)=0$ throughout the admissible solution, i.e., $x(t)\in\mathcal X^*$.

Along the solution, $\Omega$ satisfies \eqref{eq:rot_dyn} with input $\tau(t)=(c_1u_t-c_2\tau_3,\,\tau_2,\,\tau_3)$, where $\tau_2,\tau_3$ are given by \eqref{eq:tau23}. By Lemma~\ref{lma:OM_bound}, there exist $\varrho\in\mathcal{KL}$ and $\varsigma\in\mathcal K$ such that
\begin{align*}
|\Omega_3(t)|\le\|\Omega(t)\| &\le \varrho\big(\|\Omega(0)\|,t\big)+\varsigma\Big(\sup_{0\le r\le t}\|\tau(r)\|\Big) \\
&\le \varrho\big(\|\Omega(0)\|,0\big)+\varsigma(\bar\tau),
\end{align*}
where $\bar\tau:=\sup_{t\in I}\|\tau(t)\|<\infty$, with $I$ the admissible solution interval.
\end{proof}

In summary, Theorem~\ref{thm:main_result} guarantees local exponential
convergence to the desired path (\ref{goal:1}) and forward invariance of the
path-following manifold under single-rotor failure (\ref{goal:2}). The tangential
input $\nu_3$ assigns the desired motion along the path, such as point
stabilization or path-speed tracking (\ref{goal:3}). Bounded applied torque yields bounded body angular velocity (\ref{goal:4}); a bound on the Euler-angle rate $\dot\psi$ additionally requires the attitude chart to remain uniformly nonsingular.

\begin{remark}\label{rem:mode}
The thrust law~\eqref{eq:ut} is singular at $\langle b_3,q\rangle=0$, so the analysis is confined to a chosen operating mode. The present results concern $\mathcal X_d$; other modes require separate admissibility and cable-tension checks.
\end{remark}

\section{Simulation}
\label{sec:simulations}
\begin{figure}[!htp]
    \centering
    \includegraphics[width=0.95\linewidth]{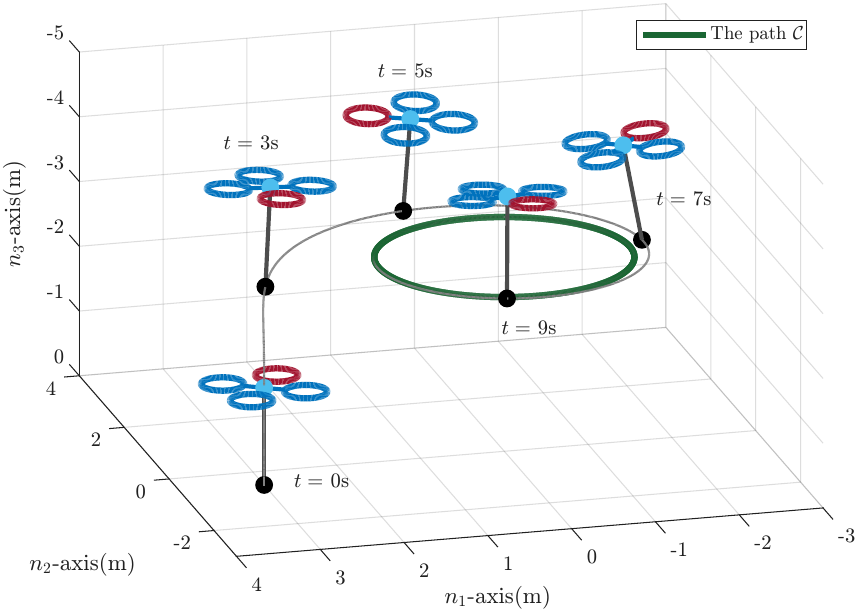}
    \caption{Snapshots of the SPS when the desired path is a horizontal circle.}
    \label{fig:snapshots}
\end{figure}

This section evaluates the proposed QSF~\eqref{eq:ut},\eqref{eq:tau23} in numerical simulations. Fig.~\ref{fig:snapshots} shows quadrotor UAV with a completely failed rotor (marked in red). Despite the failure, the payload position converges to a desired path, which is chosen to be a circle of radius $d^\ast$ at a height $r^\ast$. Hence, we pick $h_1 = p_1^2+p_2^2-(d^\ast)^2$, $h_2 = p_3- r^\ast$. The payload  moves with constant speed $v_d$ as it traverses the circle. Thus, we pick $h_3 = d^\ast\operatorname{atan2}(p_2,p_1)$. This angular path coordinate is used locally; a complete circle requires consistent treatment of the angle branch. Here, we enforce a constant arc-length velocity, i.e., $\dot h_3 \to v_d$ (see Remark~\ref{rmk:tangential_control}). We set $d^\ast=\SI{1.5}{\meter}$, $r^\ast=\SI{-3}{\meter}$, and $v_d=\SI{1}{\meter\per\second}$. The UAV parameters are $m_q = \SI{1.8}{\kilogram}$, $J = \mathrm{diag}(0.03,0.03,0.09)\,\unit{\kilogram\cdot\meter\squared}$, rotor-to-CoM distance $\ell = \SI{0.5}{\meter}$. The drag-to-thrust ratio is $c_d = 0.01$, and the damping is $k_\Omega = \text{diag}( 2.4389, 2.4389,2.4389)$. The pendulum length is $L=\SI{1.5}{\meter}$ and the payload mass is $\SI{0.4}{\kilogram}$. The control gains are obtained by placing the eigenvalues of the $z_1^\pitchfork$, $z_2^\pitchfork$, and path-speed error subsystems at $\{-1.6, -1.5, -1.4, -1.3, -1.2, -1.1\}$, $\{-1,-2\}$, and $\{-1.5, -1.4, -1.3, -1.2, -1.1\}$, respectively, which yields $K_1=[5.7658,   26.0474,   48.8674,   48.735,   27.25,    8.1]$, $K_2=[2, 3]$, and $K_3=[ 3.6036,   14.0274,   21.775,   16.85,    6.5]$. The SPS is initialized at rest with UAV horizontal and pendulum pointing down. The payload is initialized at  $p(0)=\col{3,-0.5,0}\,\unit{\meter}$.

\begin{figure}[!htp]
    \centering
    \includegraphics[width=0.95\linewidth]{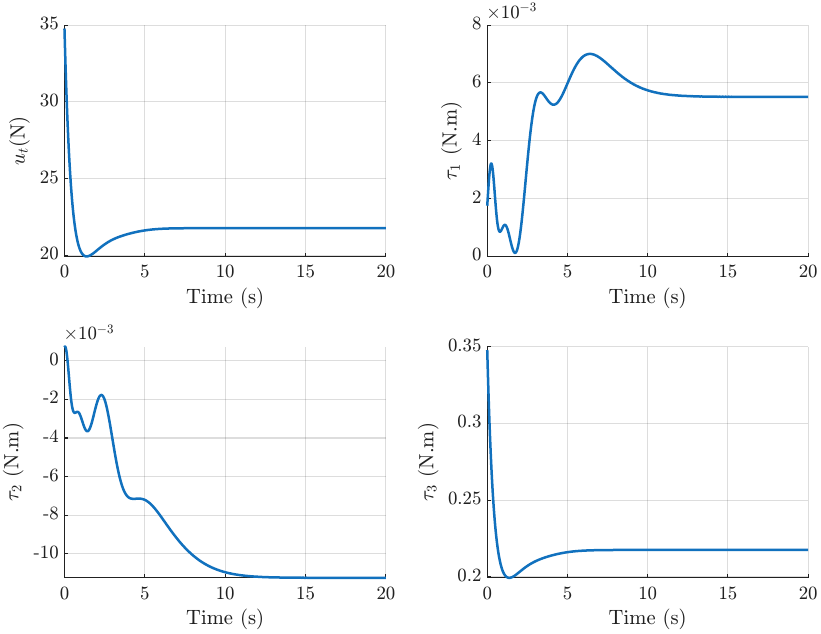}
    \caption{Inputs that generate the motion in Fig.~\ref{fig:snapshots}.}
    \label{fig:input}
\end{figure}

\begin{figure}[!htp]
    \centering
    \includegraphics[width=0.95\linewidth]{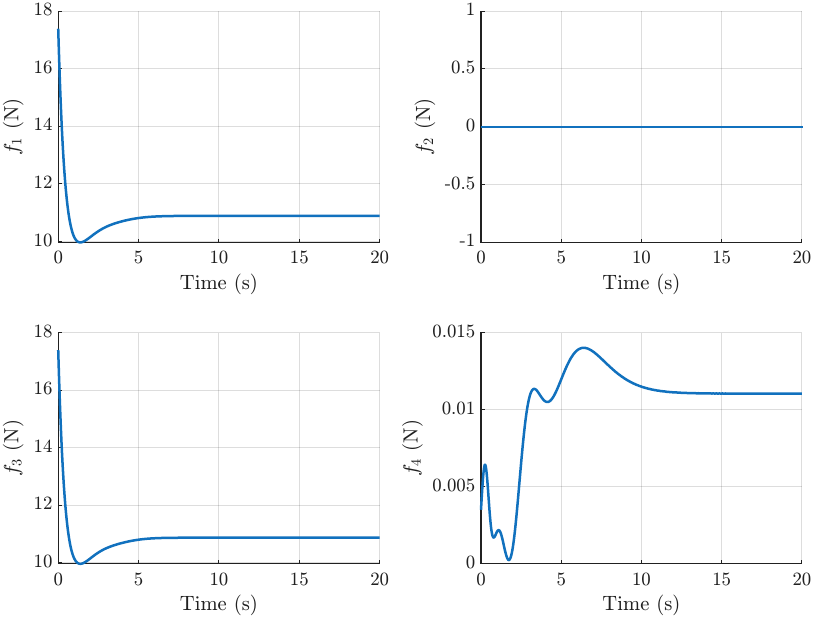}
    \caption{Rotor thrust forces generated by input in Fig.~\ref{fig:input}.}
    \label{fig:F}
\end{figure}

\begin{figure}[!htp]
    \centering
    \includegraphics[width=0.95\linewidth]{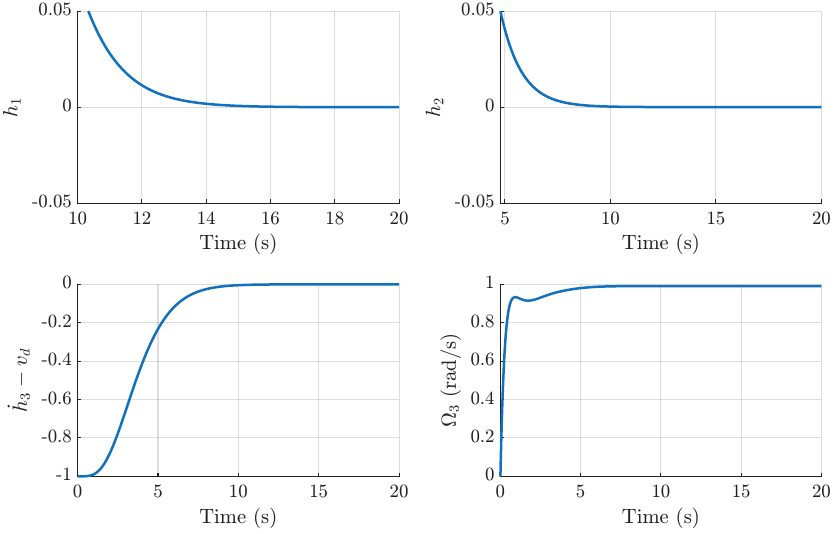}
    \caption{SPS Output and heading rate of the motion shown in Fig.~\ref{fig:snapshots}.}
    \label{fig:output}
\end{figure}

Fig.~\ref{fig:input} and Fig.~\ref{fig:output} show the input, output, and heading rate time plots for the motion depicted in Fig.~\ref{fig:snapshots}. The failure makes $\tau_1$ dependent on $u_t$ and $\tau_3$. The physical thrust forces $f_i$, $i=1,2,3,4$, generated by each rotor, are shown in Fig.~\ref{fig:F}. The outputs $h_1$, $h_2$ exponentially decay to zero, but since we control payload velocity, $\dot h_3 - v_d$ vanishes exponentially. The UAV heading rate converges to approximately $\SI{0.99}{\radian\per\second}$ (see Fig.~\ref{fig:output}).

\begin{figure}[!htp]
    \centering
    \includegraphics[width=0.95\linewidth]{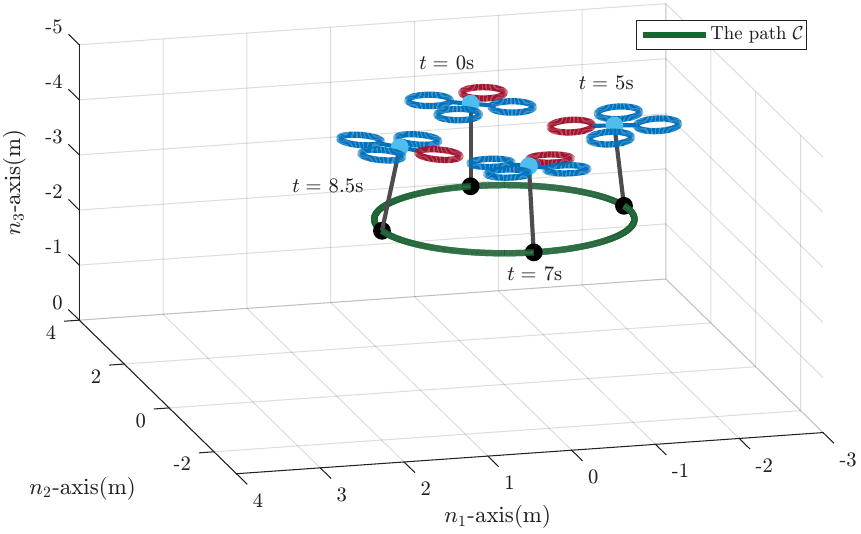}
    \caption{SPS snapshots illustrating controlled-invariance.}
    \label{fig:snapshot_inv}
\end{figure}

\begin{figure}[!htp]
    \centering
    \includegraphics[width=0.95\linewidth]{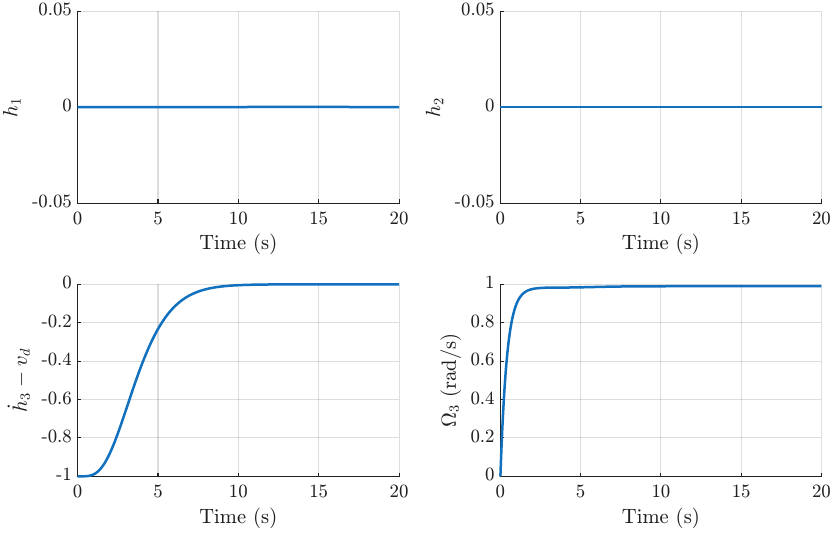}
    \caption{SPS output and heading rate for the motion depicted in Fig.~\ref{fig:snapshot_inv}.}
    \label{fig:output_inv}
\end{figure}

To validate controlled-invariance of the closed loop, we initialize the SPS on the path-following manifold, i.e., $T^\pitchfork(x(0))=0$, with the payload initially at rest on the desired path $\cC$. Fig.~\ref{fig:snapshot_inv} shows the resulting motion. The output is shown in Fig.~\ref{fig:output_inv} indicating that $h_1=h_2=0$ for all time. This means that the payload does not leave the path $\cC$ during transients. The UAV heading rate converges to approximately $\SI{0.99}{\radian\per\second}$ (Fig.~\ref{fig:output_inv}).

\begin{figure}[!htbp]
    \centering
    \begin{subfigure}[t]{0.48\columnwidth}
        \centering
        \includegraphics[width=\textwidth]{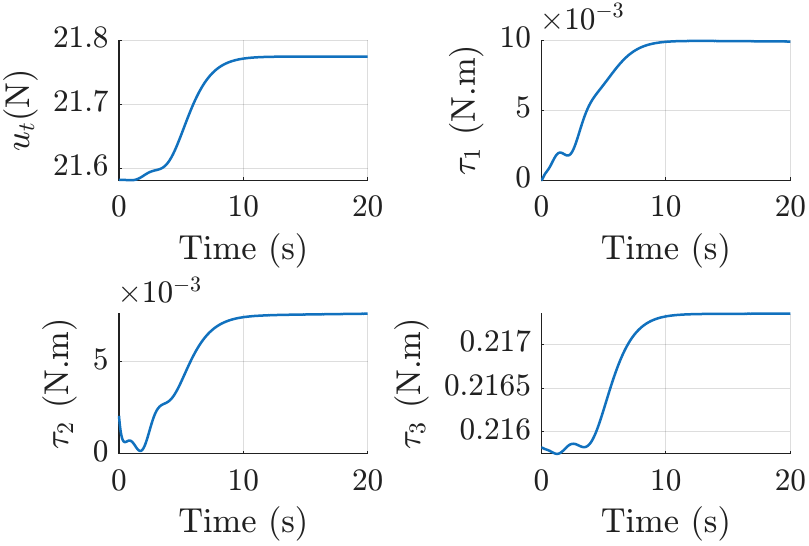}
        \caption{Thrust $u_t$ and body torques~$\tau$.}
        \label{fig:uttau}
    \end{subfigure}
    \hfill
    \begin{subfigure}[t]{0.48\columnwidth}
        \centering
        \includegraphics[width=\textwidth]{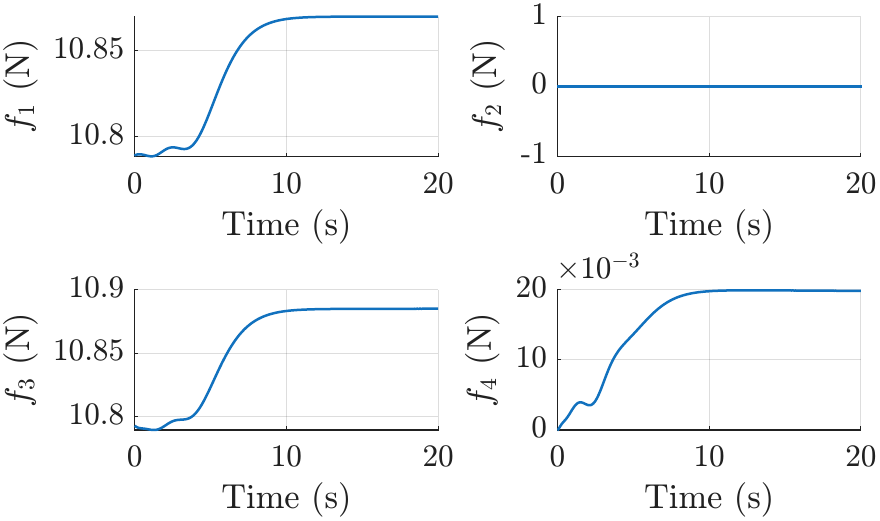}
        \caption{Individual propeller thrusts~$f_i$.}
        \label{fig:fs}
    \end{subfigure}
    \caption{SPS input for the motion depicted in Fig.~\ref{fig:snapshot_inv}.}
    \label{fig:input_inv}
\end{figure}

\section{Conclusions}\label{sec:conc}
This paper presented a quasi-static feedback (QSF) law for path following of a quadrotor carrying a slung payload under complete single-rotor failure. The faulty system was modeled directly in control-affine form, and a QSF law was designed to decouple the dynamics into transverse, tangential, and internal states. The proposed controller exponentially stabilizes the transverse dynamics, renders the path-following manifold controlled invariant, and allows independent assignment of tangential motion, including exponential path-speed tracking. Furthermore, bounded UAV body angular velocity is established under bounded applied torque. Numerical simulations using the coordinate-free SPS model demonstrate the proposed controller's performance.

\textit{Generative AI Disclosure:} ChatGPT~\cite{openai_chatgpt_2026} has been used to improve the syntax and grammar of some paragraphs.

\addtolength{\textheight}{-2cm}   %

\printbibliography

\end{document}